\documentclass[letterpaper]{article} 
\usepackage{aaai24}  
\usepackage{times}  
\usepackage{helvet}  
\usepackage{courier}  

\usepackage[hyphens]{url}  
\usepackage{arydshln}
\usepackage{graphicx} 
\usepackage{natbib}  
\usepackage{caption} 
\usepackage{algorithm}
\usepackage{algorithmic}
\usepackage{graphicx}
\usepackage{amsmath}
\usepackage{multirow}
\usepackage{amsfonts}
\usepackage{csquotes}
\usepackage[dvipsnames]{xcolor}
\usepackage[most]{tcolorbox}
\usepackage{tcolorbox}
\usepackage{subcaption}
\usepackage{lipsum}

\usepackage{todonotes}
\usepackage{newfloat}
\usepackage{listings}
\DeclareCaptionStyle{ruled}{labelfont=normalfont,labelsep=colon,strut=off} 
\floatstyle{ruled}
\newfloat{listing}{tb}{lst}{}
\floatname{listing}{Listing}
\title{Revisiting Document-Level Relation Extraction \\ with Context-Guided Link Prediction}
\author {
    Monika Jain\textsuperscript{\rm 1},
    Raghava Mutharaju\textsuperscript{\rm 1},
    Ramakanth Kavuluru\textsuperscript{\rm 2},
    Kuldeep Singh\textsuperscript{\rm 3}
}
\affiliations {
    \textsuperscript{\rm 1}Indraprastha Institute of Information Technology, Delhi, India\\
    \textsuperscript{\rm 2}University of Kentucky, Lexington, Kentucky, United States  \\
    \textsuperscript{\rm 3}Cerence GmbH and Zerotha Research, Germany \\
   \{monikaja, raghava.mutharaju@iiitd.ac.in\},
    ramakanth.kavuluru@uky.edu,
    kuldeep.singh1@cerence.com
}

\usepackage{bibentry}

\begin{document}

\maketitle

\begin{abstract}

Document-level relation extraction (DocRE) poses the challenge of identifying relationships between entities within a document as opposed to the traditional RE setting where a single sentence is input. Existing approaches rely on logical reasoning or contextual cues from entities. This paper reframes document-level RE as link prediction over a knowledge graph with distinct benefits: 1) Our approach combines entity context with document-derived logical reasoning, enhancing link prediction quality. 2) Predicted links between entities offer interpretability, elucidating employed reasoning. 
We evaluate our approach on three benchmark datasets: DocRED, ReDocRED, and DWIE. The results indicate that our proposed method outperforms the state-of-the-art models and suggests that incorporating context-based link prediction techniques can enhance the performance of document-level relation extraction models. 


\end{abstract}

\section{Introduction}
Relation extraction is the task of extracting semantic links or connections between entities from an input text~\cite{baldini-soares-etal-2019-matching}.
In recent years, document-level relation extraction problem~(DocRE) evolved as a new subtopic due to the widespread use of relational knowledge in knowledge graphs~\cite{yu-etal-2017-improved} and the inherent manifestation of cross-sentence relations involving multi-hop reasoning. 
Thus, as compared to traditional RE, DocRE has two major challenges: subject and object entities in a given triple might be dispersed across distinct sentences, and certain entities may have aliases in the form of distinct entity mentions. Consequently, the signal (hints) needed for DocRE is not confined to a single sentence. A common approach to solve this problem is by taking the input sentences and constructing a structured graph based on syntactic trees, co-references, or heuristics to represent relation information between all entity pairs~\cite{nan-etal-2020-reasoning}. A graph neural network model is applied to the constructed graph, which performs multi-hop graph convolutions to derive features of the involved entities. A classifier uses these features to make predictions~\cite{DBLP:conf/aaai/ZhangCZW20}. Another approach ~\cite{xu-etal-2021-discriminative,10.1145/3572898} explicitly models the reasoning process of different reasoning skills (e.g., multi-hop, coreference-mediated). However, even after considering features between the entity pairs and executing the reasoning process, DocRE is still hard due to the latent and unspecific/imprecise contexts. 

\begin{figure}[ht]
    \centering
    \includegraphics[width=0.45\textwidth]{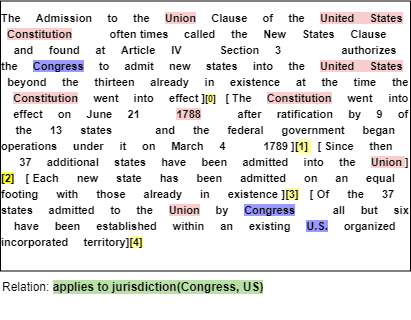}
    \caption{A partial document and labeled relation from DocRED. Blue color represents concerned entities, pink color represents other mentioned entities, and yellow color denotes the sentence number.}
    \label{example}
\end{figure}

Consider the sentence in Figure~\ref{example} and its labeled relation \textit{applies\_to\_jurisdiction} (Congress, US) from the DocRED dataset~\cite{yao-etal-2019-docred}. Even with the inclusion of multi-hop and co-reference reasoning, inferring the correct relation becomes challenging because the relation depends on multiple sentences and cannot be identified based on the language used in the sentence.

For these kinds of sentences, external context (knowledge) can play a vital role in helping the model capture more about the involved entities. For the above example, using the Wikidata knowledge base~\cite{10.1145/2629489} and WordNet~\cite{10.1145/219717.219748}, we can get details, such as the entity types, synonyms,  and other direct and indirect relations between entities (if they exist) on the Web. 
For these kinds of sentences, external context (knowledge) can play a vital role in helping the model capture more about the involved entities. For the above example, using the Wikidata knowledge base~\cite{10.1145/2629489}, we can get additional details, such as the entity types, synonyms,  and other direct and indirect relations between entities (if they exist) on the Web. 

Previous research in this domain has underscored the potential of external context to enhance performance in relation extraction, co-reference resolution, and named entity recognition~\cite{shimorina-etal-2022-knowledge}.
The distinctive innovation of our work lies in the fusion of context extracted from Wikidata and WordNet with a reasoning framework, enabling the prediction of entity relationships based on input document observations. Given the wide availability of external context, Knowledge Graph (KG) triples can augment training data, thereby casting the DocRE task as a knowledge Graph based link prediction challenge. In other words, given head and tail entities, we address the question of determining the appropriate relation.

We demonstrate that framing DocRE as a link prediction problem, combined with contextual knowledge and reasoning, yields enhanced accuracy in predicting the relation between entities. We furnish traversal paths as compelling justifications for relation predictions, thereby shedding light on why a particular relation is favored over others. Notably, this marks the first instance of presenting a traversal path between entities for each prediction in the context of DocRE. Our contributions in this work are as follows.

\begin{itemize}
  \item We introduce an innovative approach named DocRE-CLiP (\textbf{Doc}ument-level \textbf{R}elation \textbf{E}xtraction with \textbf{C}ontext-guided \textbf{Li}nk \textbf{P}rediction), which amalgamates external entity context with reasoning via a link prediction algorithm.
  \item  Our empirical analyses encompass three widely-used public document-level relation extraction datasets, showcasing our model's improvement over the recent state-of-the-art methods.
  \item Significantly, for every prediction, our approach is first in DocRE literature to supply a traversal path as corroborative evidence, bolstering the model's interpretability.
\end{itemize}


\section{Related work}
\textbf{DocRE} Previous efforts~\cite{zhang-etal-2018-graph, 10.1007/978-3-031-43421-1_14} in relation extraction have focused on predicting relationships within a single sentence. Recent years have seen growing interest in relation extraction beyond single sentence \cite{yao-etal-2019-docred}.

\textbf{Transformer-Based DocRE} is another interesting approach to tackle the document-level relation extraction problem~\cite{zeng-etal-2020-double}. One primary focus revolves around maximizing the effective utilization of long-distance token dependencies using a transformer. 
Earlier research considers DocRE as semantic segmentation task, employing the entity matrix, and they utilize a U-Net to capture and model~\cite{ijcai2021p551}. 
In a separate study, localized contextual pooling was introduced to focus on tokens relevant to individual entity pairs ~\cite{DBLP:conf/aaai/Zhou0M021}.
On the other hand, the DocRE challenge is addressed by incorporating explicit supervision for token dependencies, achieved by leveraging evidential information~\cite{DBLP:conf/eacl/MaWO23}.  

\textbf{Graph-Based DocRE} is based on a graph constructed with mentions, entities, sentences, or documents. The relations between these nodes are then deduced through reasoning on this constructed graph. Earlier research in this line of work solves DocRE with multi-hop reasoning on a mention level graph for inter-sentential entity pair \cite{zeng-etal-2020-double}. 
A discriminative reasoning framework (DRN) is introduced in a different study. This framework involves modeling the pathways of reasoning skills connecting various pairs of entities~\cite{xu-etal-2021-discriminative}. DRN is designed to estimate the relation probability distribution of different reasoning paths based on the constructed graph and vectorized document contexts for each entity pair, thereby recognizing their relation. We have used DRN as a base model for implementing reasoning skills. 

\textbf{Context Knowledge Based RE} study the context integration with model primarily through knowledge base (KB). Past work in this line of work uses entity types and entity aliases to predict the relation~\cite{vashishth-etal-2018-reside,10.1007/978-3-030-62419-4_11}. RECON~\cite{DBLP:conf/www/BastosN0MSHK21} encoded attribute and relation triples in the Knowledge Graph and combined the embedding with their corresponding sentence embedding. KB-Both~\cite{DBLP:conf/acl/VerlindenZDDD21} uses entity details from hyperlinked text documents from Wikipedia and Knowledge Graph (KG) from Wikidata to enhance performance.
In distinction to that~\cite{DBLP:conf/semweb/WangWSH22} integrate knowledge, including co-references, attributes, and relations with different injection methods, for improving the state-of-the-art. In contrast to these approaches, we consider the context of entities and the external relation paths between them. Furthermore, we employ a reasoning model that effectively addresses DocRE and enhances the robustness of the proposed approach.

\section{Methodology}
\subsection{Problem Formulation}
An unstructured document D consisting of K sentences is represented by $\{S\}_{i=1}^{k}$, where each sentence is a sequence of words and entities $\mathcal{E}$=$\{{e_i}\}_{i=1}^{P}$(P is the total number of entities). Entities $e_{i}$  has multiple mentions, $m_{i}^{s_{k}}$, scattered across the document D. Entity alias are represented as $\{{m_{i}^{s_{k}}}\}_{k=1}^{Q}$. Our objective is to extract the relation between two entities in $\mathcal{E}$ namely P(r$\mid${e$_i$,e$_j$}) where $e_i,e_j\in \mathcal{E}, r\in \mathcal{R}$, here $\mathcal{R}$ is a total labeled relation set. The context (background knowledge) of an entity ${e_i}$ is represented by $C_{e_i}$ and a context path, i.e., a sequence of connected entities and edges from the head entity (e$_i$) to the tail entity (e$_j$) is represented by $CP_{e_i,e_j}$.

\subsection{Approach}
\begin{figure*}[ht]
    \centering
   \includegraphics[width=0.7\textwidth]{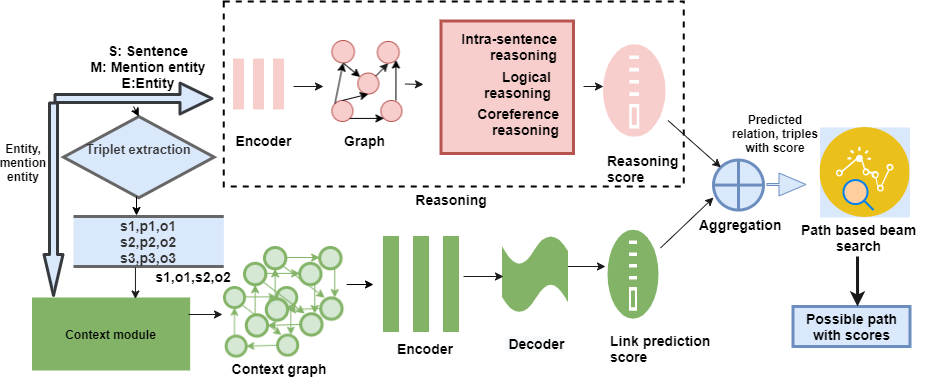}
    \caption{ Illustration of proposed framework DocRE-CLiP and its various modules.}
    \label{DocRE-CLiP}
\end{figure*}

 Our proposed framework, DocRE-CLiP, integrates document-derived reasoning with context knowledge using link prediction. In the first step, we extract triples from the sentences of the given document. In the second step, we extract two types of context: 1) entity context, such as its aliases, and 2) context paths from an external KB  (Wikidata, in this case). Using the triples and extracted contexts, we create a context graph to calculate a link prediction score. Then in the third step, we use several reasoning mechanisms such as logical reasoning, intra-sentence reasoning, and co-reference reasoning to calculate relation scores for pairs of entities. In the final step,  
the aggregation module combines the relation scores from the second and third steps. We have also implemented a path-based beam search in the framework to explain the predicted relation by providing traversal paths based on scores (refer to Figure \ref{DocRE-CLiP}). We now detail the architecture of our proposed framework.

\paragraph{
\textbf{Triplet Extraction Module.}} Document Relation Extraction (DocRE) datasets often contain labeled triplets; however, popular datasets~\cite{yao-etal-2019-docred} have about 64.6\% missing triples, yielding an incomplete graph~\cite{DBLP:conf/emnlp/Tan0BNA22}. To extract all the triples from the  document, we utilize an open-source state-of-the-art method~\cite{huguet-cabot-navigli-2021-rebel-relation} for triplet generation. This model is chosen based on source code availability and run time. This module takes a document $D$ as input and produces triples ($s$, $p$, $o$), where $s$ is the head entity, $p$ is the relation, and $o$ is the tail entity. These extracted triples ($T$) follow the equation below, where $n$ is the total count of triples extracted from document $D$.
\begin{equation}\label{eq3}
    T(D)=\{s_i,p_i,o_i\}_{i=1}^n
\end{equation}

\paragraph{\textbf{Context Module.}} Our goal involves extracting two types of contexts -- entity context and the contextual path between entity pairs. For entity $e_i$, we generate entity context using entity type and synonyms, which are derived by using WordNet~\cite{10.1145/219717.219748}.
We incorporate the entity context $C_{e_i}$ into the triples. Here, $S$ denotes the total number of extracted synonyms. Refer to equations \ref{synonym} and \ref{type} for more details.

 \begin{equation}\label{synonym}
C_{e_i}=\{e_i, hasSynonym, Synonym_k\}_{k=1}^S, 
\end{equation}
\begin{equation}\label{type}
C_{e_i}=\{ e_i, hasEntityType,{EntityType}\} 
\end{equation}



Let us consider an entity \enquote{USA} as an example. When we utilize WordNet, we discover that synonyms for \enquote{US} include, \enquote{America}, and \enquote{United States}. Additionally, the entity is categorized as the type \enquote{Country}. Consequently, the following set of triples is generated through this process: 
\begin{tcolorbox}
\centering
\small
\vspace{-0.1cm}
\textit{\{{USA, hasSynonym, US}\} }\\
\textit{\{{USA, hasSynonym, America}\} }\\
\textit{\{{USA, hasSynonym, United States}\}} \\
 \textit{\{{USA, hasEntityType, Country}\}} \\
  \vspace{-0.1cm}  
\end{tcolorbox}


The second source of contextual information pertains to entity paths. Predicting relations between entity pairs poses challenges stemming from inherent document deficiencies~\cite{DBLP:conf/emnlp/Tan0BNA22}. To address these issues, we introduced external context by harnessing insights from Wikidata.
The procedure involves extracting paths (direct and indirect) between entity-entity, mention-entity, and entity-mention pairs from Wikidata, provided they exist. The contextual path pertains to an entity pair ${{e_i,e_j}}$. We considered context paths spanning an N-hop distance (N being chosen based on experimental findings) between the entity pair. Subsequently, the extracted path is transformed into triples and forwarded to the link prediction model.

\begin{figure}[ht]
    \centering
    \includegraphics[width=0.5\textwidth]{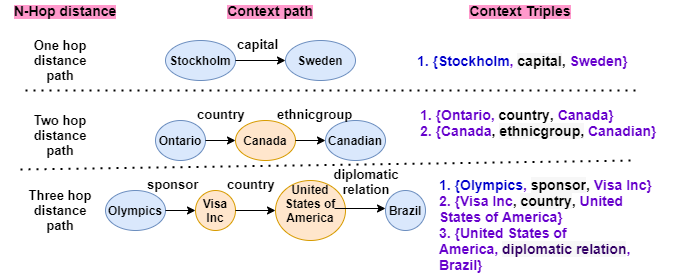}
    \caption{Triples constructed using N-hop path extracted from Wikidata. The head and tail entities are blue in color.  Intermediate entities are in peach color.}
    \label{hopdistance}
\end{figure}

Illustrated in Figure~\ref{hopdistance} is the triple generation process employing a contextual path. The entity \textit{Canadian} is two hops away from the entity \textit{Ontario}. \textit{Canada} is an intermediary entity, while \textit{country} and \textit{ethnic group} are intermediary properties. The Contextual Path  ($CP_{e_i,e_j}$) are a set of triples formed using the intermediary entities and properties, as shown in Figure~\ref{hopdistance}.



\paragraph{\textbf{Link Prediction Module.}} Link prediction is the task of predicting absent or potential connections among nodes within a network~\cite{10.1145/956863.956972}. Given that the document relation extraction (DocRE) task involves constructing a graph that interlinks entities and considering our context, formulated as triples, which can be conceptualized as a Knowledge Graph (KG), we approach the DocRE challenge as a link prediction problem. This approach encompasses both an encoder and a decoder. The encoder maps each entity
\( e_{i} \in \mathcal{E} \) to a real-valued vector $v_i \in  \mathbb{R}^{d}$, where $\mathbb{R}$ denotes the set of real-valued relation vectors of dimension $d$.
The decoder reconstructs graph edges by leveraging vertex representations, essentially scoring (subject, relation, object) triples using a function:
\( \mathbb{R}^{d} \times \mathcal{R} \times \mathbb{R}^{d} \rightarrow \mathbb{R} \).
While prior methods often employ a solitary real-valued vector $e_i$ $\in \mathcal{E}$, our approach computes representations using an R-GCN \cite{DBLP:conf/esws/SchlichtkrullKB18} encoder, where $h_i^{(l)}$ is the hidden state of node $e_i$ in the l-th layer of the neural network. To compute the forward pass for an entity $e_i$ in a relational multi-graph, the propagation model at layer $l+1$ is computed as follows.
\begin{equation}
h_i^{(l+1)}=\sigma \left(\sum_{r\in{R}}\sum_{j\in{N_i^{r}}}(\frac{1}{c_{i,r}}W_r^{(l)}h_j^{(l)}+W_0^{(l)}h_{i}^{(l)}) \right)
\end{equation}

Here, $h_{i}^{(l)}$ $\in \mathbb{R}^{d^{(l)}}$ with $d^{(l)}$ being the dimensionality of layer $l$. $W_0^{l}$ and $W_r^{(l)}$ represent the block diagonal weight matrices of the neural network, and $\sigma$ represents the activation function. $N_i^{r}$ signifies the set of neighboring indices of node $i$ under relation $r \in R$, and $c_{i,r}$ is a normalization constant.

For training the link prediction model, our dataset comprises a) core training triples from the dataset, b) triplets obtained through the triplet extraction module using Equation \ref{eq3}, c) triplets formulated using the context module guided by Equations \ref{synonym} and \ref{type}, and d) triplets constructed using context paths connecting entity pairs.
We use DistMult~\cite{DBLP:journals/corr/YangYHGD14a} as the decoder. It performs well on the standard link prediction benchmarks. Every relation r in a triple is scored using equation \ref{eq8}.

\begin{equation}\label{eq8}
    P(r \mid i,j)=P(e_i^T \times R_r \times e_j)
\end{equation}


\paragraph{\textbf{Reasoning Module.}} We consider three types of reasoning in our approach. \\
  1) \textit{Intra-sentence reasoning}, which is a combination of pattern recognition and common sense reasoning. 
  Intra-sentence reasoning path is defined as \( P I_{i j}=m_{i}^{s_{1}} \circ s_{1} \circ m_{j}^{s_{1}} \) for entity pair \{$e_i,e_j$\} insider the sentence $s_1$ in document D. $m_i^{s_1}$ and $m_j^{s_1}$ are mentions and \enquote{o} denotes reasoning step on reasoning path from $e_i$ to $e_j$.  \\
  2) \textit{Logical reasoning} is where a bridge entity indirectly establishes the relations between two entities.
  Logical reasoning path is formally denoted as \( P L_{i j}=m_{i}^{s_{1}} \circ s_{1} \circ m_{l}^{s_{1}} \circ m_{l}^{s_{2}} \circ \ s_{2}\  \circ \)  \(\  m_{j}^{s_{2}} \) for entity pair \{$e_i,e_j$\} from sentence $s_1$ and $s_2$ is directly established by bridge entity $e_l$.   \\
  3) \textit{Co-reference reasoning} which is nothing but co-reference resolution. 
  Co-reference reasoning path is defined as \( PC_{ij}=m_{i}^{s_{1}} \circ s_{1} \circ s_{2} \circ m_{j}^{s_{2}} \) between two entities $e_i$ and $e_j$ which occur in same sentence as other entity. Our implementation of these reasoning skills is inspired by ~\cite{xu-etal-2021-discriminative}.

Consider an entity pair~$\{e_i,e_j\}$ and its intra sentence reasoning path (PI$_{i j}$), logical reasoning path (PL$_{i j}$) and co-reference reasoning path (PC$_{i j}$) in the sentence. The various reasoning is modeled to recognize the entity pair as intra-sentence reasoning \( \mathrm{R}_{P I}(r)=P\left(r \mid e_{i}, e_{j}, P I_{i j}, D\right) \), logical reasoning \( \mathrm{R}_{P L}(r)=P\left(r \mid e_{i}, e_{j}, P L_{i j}, D\right) \)
 and co-reference reasoning \( \mathrm{R}_{P C}(r)=P\left(r \mid e_{i}, e_{j}, P C_{i j}, D\right) \).
 Reasoning type is selected with max probability to recognize the relation between each entity pair using the equation:
 \begin{equation}
     P\left(r \mid e_{i}, e_{j}, D\right)=\max \left[\mathrm{R}_{P I}(r), \mathbf{R}_{P L}(r), \mathbf{R}_{P C}(r)\right] 
 \end{equation}

For discerning relations between two entities, we employ two categories of context representation -- heterogeneous graph context representation (HGC) and document-level context representation (DLC) to model diverse reasoning paths~\cite{DBLP:conf/aaai/Zhou0M021}.
In heterogeneous graph context representation (HGC), a word is portrayed as a concatenation embedding of its word ($W_e$), entity type ($W_t$), and co-reference embedding ($W_c$). This composite embedding is then input into a BiLSTM to convert the document $D$ into a vectorized form using the equation: BiLSTM([W$_e$:W$_t$:W$_c$]). Following the methodology of \cite{zeng-etal-2015-distant}, a heterogeneous graph is constructed based on sentence and mention nodes.
For document-level context representation, following~\cite{DBLP:conf/icra/EisenbachLAG23}, a self-attention mechanism is employed to learn document-level context (DLC) for a specific mention based on the vectorized input document $D$.


To model intra-sentence reasoning path (\( \alpha_{i j})\), logical reasoning path (\(\beta_{i j} \)) and co-reference reasoning path (\(\gamma_{i j} \)), HGC and DLC representation are combined~\cite{DBLP:conf/aaai/XuCZ21}.
These reasoning representations are the input to the classifier to compute the
probabilities of the relation between $e_i$ and $e_j$ entities by a multi-layer perceptron (MLP) for each path, respectively (equation~\ref{eq2}).

\begin{equation}\label{eq2}
    P\left(r \mid e_{i}, e_{j}\right) = \max\left[
    \begin{aligned}
        &\operatorname{sigmoid}\left(\operatorname{MLP}_{r}\left(\alpha_{ij}\right)\right), \\
        &\operatorname{sigmoid}\left(\operatorname{MLP}_{r}\left(\beta_{ij}\right)\right), \\
        &\operatorname{sigmoid}\left(\operatorname{MLP}_{r}\left(\gamma_{ij}\right)\right)
    \end{aligned}
    \right]
\end{equation}

By the end of this step, we will get the score of each relation ($r$) for a given $\{e_i,e_j\}$.

\paragraph{\textbf{Aggregation Module.}} In this module, we aggregate the probability score from the reasoning module equation~\ref{eq2} and link prediction probability score using equation~\ref{eq8}. Further, the binary cross-entropy is used as a training objective function~\cite{yao-etal-2019-docred} for predicting the final relation.

\subsection{Path-Based Beam Search}
An essential component of our approach is that it can explain the predicted relation by providing the most relevant path in the graph between the given entity pairs. This represents a notable advancement, as contemporary state-of-the-art models cannot often furnish explanations alongside their predictions. Unlike Greedy search, where each position is assessed in isolation, and the best choice is selected without considering preceding positions, we use beam search. This strategy selects the top \enquote{N} sequences thus far and factors in probabilities involving the concatenation of all previous paths and the path in the current position.

Inspired by~\cite{10.1145/3514221.3517887}, we used beam search to derive plausible paths leading to the target entity within a graph. This graph (G) is constructed using the triples to train the link prediction module, augmented by test result triplets from the model's predictions. Our objective is to create a comprehensive graph encompassing the maximum available details, to generate substantial explanations for the predictions.
Formally, we conceptualize the path-based beam search challenge as follows. Given a structured relational query ($e_i$, $r$, ?), where $e_i$ serves as the head entity, $r$ signifies the query relation, and ($e_i$, $r$, $e_j$) $\in$ G, our objective is to identify a collection of plausible answer entities {e$_j$} by navigating paths through the existing entities and relations within G, leading to tail entities. We compile a list of distinct entities reached during the final search step and assign the highest score attained among all paths leading to each entity. Subsequently, we present the top-ranked unique entities. This approach surpasses the direct output of entities ranked at the beam's apex, which often includes duplicates. Upon completing this step, we obtain actual paths (sequences of nodes and edges) for enhanced interpretability.
\section{Experimental Setup}
We conduct our evaluation in response to the following research questions.
\textbf{RQ1}: What is the effectiveness of DocRE-CLiP that combines context knowledge with reasoning in solving document-level relation extraction tasks? \textbf{RQ2}: How does knowledge encoded from external sources impact the performance of DocRE-CLiP? \textbf{RQ3}: Does the explanation generated by our approach provide sufficient grounds to support the inferred relation?
\subsection{Datasets}
The proposed model is evaluated on three widely-used public datasets 1) DocRED~\cite{yao-etal-2019-docred} 2) ReDocRED~\cite{DBLP:conf/eacl/MaWO23} and 3) DWIE~\cite{DBLP:journals/ipm/ZaporojetsDDD21}. ReDocRED is a revised version of handling DocRED issues such as false negatives and incompleteness~\cite{DBLP:conf/emnlp/Tan0BNA22}. Dataset details are in Table~\ref{dataset}.
\begin{table}[ht] 
 \scalebox{0.80}{
    \begin{tabular}{cccccc}
\hline
\textbf{Dataset} & \textbf{\#Triples} & \textbf{\#Rel} & \textbf{\#Entities} & \textbf{\#Entity Types}& \textbf{\#Doc}   \\
\hline
 DocRED & 50,503 & 96 & 30554 & 7 & 5053  \\ 

\hline
ReDocRED & 120,664 & 96& 38239 &  10 & 5053 \\ 
\hline
DWIE & 19465 & 66 & 6644 & 10 &777 \\ 
\hline
\end{tabular}
}
\caption{Dataset statistics}
    \label{dataset}   
\end{table}

\subsection{Baseline Models for Comparison}
We used several competitive baselines and a recent state-of-the-art dataset for comparison. For DocRED, we compared our approach with  
BERT based models such as 
SIRE~\cite{zeng-etal-2021-sire},
HeterGSAN-Rec~\cite{DBLP:conf/aaai/XuCZ21},
ATLOP~\cite{DBLP:conf/aaai/Zhou0M021}, 
DRN~\cite{xu-etal-2021-discriminative}, and RoBERTa based model such as DREEAM~\cite{DBLP:conf/eacl/MaWO23}. 
and ATLOP~\cite{DBLP:conf/aaai/Zhou0M021}, 
KD-DocRE~\cite{tan-etal-2022-document}, 
DocuNet~\cite{ijcai2021p551}, 
EIDER~\cite{xie-etal-2022-eider}, SAIS~\cite{xiao-etal-2022-sais},
~\cite{DBLP:conf/aaai/Zhou0M021} are also compared with the proposed DocRE-CLiP. 

Similarly, for the ReDocRED dataset, we used 
ATLOP~\cite{DBLP:conf/aaai/Zhou0M021},
DRN~\cite{xu-etal-2021-discriminative},  
DocuNet~\cite{ijcai2021p551}, 
KD-DocRE~\cite{tan-etal-2022-document} and the best baseline
DREEAM$_{inference}$~\cite{DBLP:conf/eacl/MaWO23}. For the DWIE dataset, we considered the state-of-the-art model 
DRN~\cite{xu-etal-2021-discriminative}. Other than baseline model, we evaluated our model with context-based models such as KIRE~\cite{DBLP:conf/semweb/WangWSH22}, RESIDE~\cite{vashishth-etal-2018-reside}, RECON~\cite{DBLP:conf/www/BastosN0MSHK21} and KB-graph~\cite{DBLP:conf/acl/VerlindenZDDD21}.

\subsection{Hyper-Parameters and Metrics}
For the reasoning module, we follow the settings of~\cite{xu-etal-2021-discriminative}. We use the word embedding from GloVe (100d) and apply a Bidirectional LSTM (128d) to a word representation for encoding. 
We employ uncased BERT-Based model (768d) as an encoder with a learning rate 1e-3. We used AdamW as an optimizer, and the learning rate is $1e-3$. 
R-GCN is used as an encoder with a single encoding layer (200d) embeddings for the link prediction model. We regularize the encoder through edge dropout applied before normalization, with a dropout rate of 0.2 for self-loops and 0.4 for other edges. We apply l2 regularization to the decoder with a penalty of 0.01. Adam~\cite{DBLP:journals/corr/KingmaB14} is used as an optimizer, and the model is trained with 100 epochs using a learning rate of 0.01. For extracting the context paths, we use SPARQL queries to retrieve paths between entities. If multiple paths exist between entities, we consider the path with the highest page rank. The N-hop path length of the context varies from 1 to 4. The rationale behind this range is that we found no pertinent information for the context beyond four hops. We have used beam size of 128 for beam search for all three datasets.

We use the evaluation metrics of DocRED~\cite{yao-etal-2019-docred}, i.e., F1 and Ign F1 for DocRE-CLiP. Ign F1 is measured by removing relations in the annotated training set from the development and test sets. 

    


\section{Results}
We have compared DocRE-CLiP with various baseline models on DocRED, ReDocRED, and DWIE datasets given in Table~\ref{results}. The results effectively address our primary research question (\textbf{RQ1}). To delve into the specifics of \textbf{(RQ1)}, we observe that incorporating context information from Wikidata and WordNet improves the performance compared to the baseline models. Notably, DocRE-CLiP surpasses all graph-based, reasoning-oriented, and transformer-based models by incorporating contextual information.

\begin{table}[ht] 
\centering
 
\scalebox{0.63}{
\begin{tabular}{cccccc}
\hline

 \textbf{Baseline} & \textbf{PLM/GNN} & \multicolumn{2}{c}{\textbf{Dev}} & \multicolumn{2}{c}{\textbf{Test}} \\

 \cline{3-4} \cline{5-6}

 {} & {} & \textbf{F1} & \textbf{Ign F1} & \textbf{F1} & \textbf{Ign F1} \\
\hline
\multicolumn{6}{c}{\textbf{Dataset-DocRED}}  \\
\hline
 SIRE & BERT & 61.6 & 59.82 & 62.05 & 60.18 \\

 HeterGSAN-Rec & BERT & 60.18 & 58.13 & 59.45 & 57.12 \\

ATLOP & BERT  & 61.09 & 59.22 & 61.30 & 59.31 \\ 
  DRN & BERT & 61.39  & 59.33 & 61.37 & 59.15 \\ 
 
 DocuNet& RoBERTa& 64.12 & 62.23  & 64.55 & 62.39 \\ 
 
 ATLOP & RoBERTa &  63.18 & 61.32 & 63.40 & 61.39 \\ 
 	
KD-DocRE & RoBERTa & 67.12 & 65.27 & 67.28 & 65.24 \\
 	
SAIS & RoBERTa  & 65.17 & 62.23 & 65.11 & 63.44 \\
 DREEAM & RoBERTa  &67.41 & 65.52 & 67.53 & 65.47 \\ 
 EIDER & RoBERTa   &64.27 & 62.34 & 64.79 & 62.85 \\ 

KIRE& - & 52.65 & 50.46 & 51.98& 49.69 \\

RESIDE& GNN & 51.59 & 49.64 & 50.71 & 48.62 \\

RECON & GNN & 52.89& 50.78 & 52.27 & 49.97  \\

KB-Graph  & -  & 52.81& 50.69 & 52.19 & 49.88\\

 {DocRE-CLiP} & BERT & {68.13$_{\pm 0.15}$}
  & {66.43$_{\pm 0.17}$} & {68.51} & {66.31}  \\ 
  
\hline

\multicolumn{6}{c}{\textbf{Dataset-DWIE}}  \\
\hline
DRN*$_{GloVe}$ & BERT  & - & -  & 56.04  & 54.22 \\

RESIDE & GNN & 65.11& 55.74 & 66.78& 57.64 \\

RECON  & GNN & 65.48& 56.12 & 66.94& 58.02\\

KB-Graph & - & 65.39 & 56.03 & 66.89 &  57.94\\

 {DocRE-CLiP} & BERT &{66.12$_{\pm 0.12}$} & {57.11$_{\pm 0.16}$} & {67.10$_{\pm 0.11}$} & {58.87$_{\pm 0.17}$} \\

\hline
\multicolumn{6}{c}{\textbf{Dataset-ReDocRED}}  \\
\hline

  ATLOP  &  BERT &- & - & 77.56 &  76.82 \\
DRN*  &  BERT & - & - & 75.6 & 74.3 \\

KD-DocRE & BERT &- & - & 81.04 &  80.32\\

DocuNet  & RoBERTa & - & - & 79.46 &  78.52 \\

 DREEAM & RoBERTa &- & - & 81.44 &  80.39\\

 {DocRE-CLiP} & BERT & - & - & {81.55$_{\pm 0.14}$} & {80.57$_{\pm 0.22}$} \\
 
\hline

\end{tabular}
}
\caption{{Results on DocRED, ReDocRED, and DWIE datasets, including the baseline models. The precision column is blank (-) for baselines that do not report it. * denotes results obtained after modifying their code as the dataset necessitates. The mean and standard deviation of F1 and IgnF1 on the dev set are reported for three training runs. We report the official test score for DocRED on the best checkpoint on the dev set.
}}
    \label{results}   
\end{table}

Examining the results for the DocRED dataset, our DocRE-CLiP model showcases an improvement of approximately 1\% compared to top-performing models like KD-DocRE and DREEAM. For the ReDocRED dataset, DocRE-CLiP outperforms baseline models, including DREEAM, KD-DocRE, and DocuNet. Furthermore, in the case of DWIE dataset, our model outperforms all the baseline models, such as DRN, GAIN, and ATLOP, KIRE. Notably, the ReDocRED dataset exhibits only slight improvement over the recent state-of-the-art DREEAM. This could be attributed to ReDocRED being an enhanced version of DocRED, which already tackled the issue of dataset incompleteness. On the other hand, both DocRED and DWIE demonstrate significant improvements by incorporating REBEL triplets, setting them apart from ReDocRED in this regard.
The unique aspect of DocRE-CLiP lies in its incorporation of contextual paths between entities in addition to entity context, contributing to its superior performance compared to context-aware approaches like KIRE, KB-both, and RESIDE, which leverage knowledge bases. This advantage is rooted in the reasoning framework that underlies DocRE-CLiP.

\section{Ablation Study}

\begin{figure}[t]
\centering
\includegraphics[width=7cm]{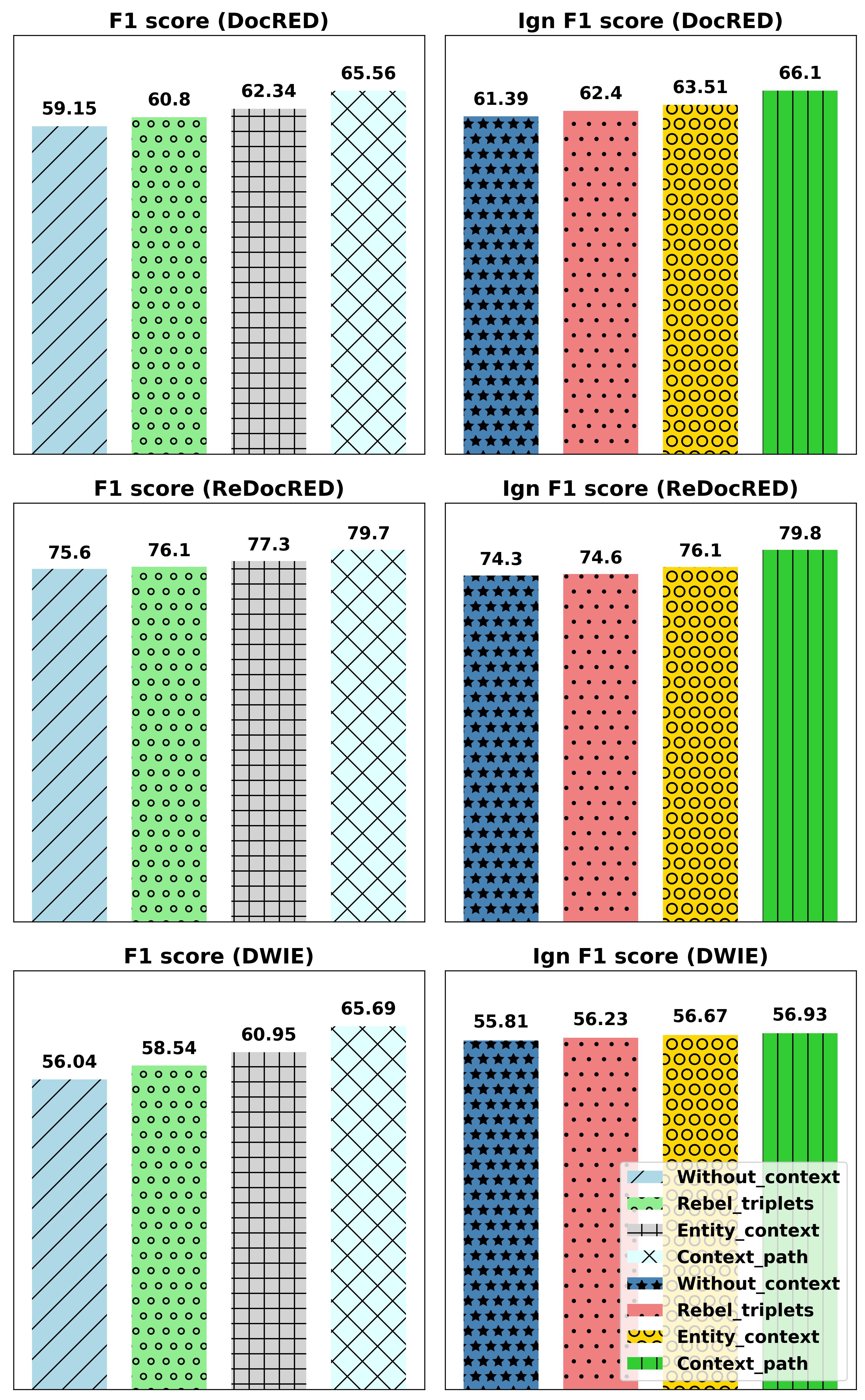}
\caption{Performance of DocRE-CLiP across various contexts using the DocRED, ReDocRED, and DWIE datasets}
    \label{contextf1score}
\end{figure}

\paragraph{\textbf{Effectiveness of different contexts on DocRE-CLiP.}} Figure~\ref{contextf1score} offers an overview of our findings into DocRE-CLiP's performance under varying context conditions. Initially, we gauged its performance without context and documented the outcomes. Subsequently, we introduced document triplets along with the dataset labels and determined the corresponding F1 scores. Additionally, we measured the impact of entity context and documented the ensuing performance. Furthermore, we scrutinized the effect of context path details on DocRE-CLiP by considering only those paths that notably enhanced performance. Our analysis establishes that the incorporation of context enhances performance across all datasets. Thus, we effectively address our second research question, \textbf{(RQ2)}.
This study's findings lead us to conclude that DocRE-CLiP benefits the most from the context path compared to the other contexts.

\begin{table}[ht]
\centering
\small 
\noindent
\scalebox{0.8}{
\begin{tabular}{@{} p{1.2\linewidth}}
\hline

 \textbf{Case1}: \\
 \hline
\textbf{Sentence 0}: The Château. de    Pirou    is    a    castle    in    the    commune    of    Pirou         in    the    département  
  of   \underline{\textbf{Manche}}    (    Normandy    )         \underline{\textbf{France}}]  \\
 \textbf{Correct answer}:  {contains\_administrative\_territorial entity} \\
 \textbf{Baseline}: country \\
 \textbf{DocRE-CLiP}: { contains\_administrative\_territorial\_entity}  \\ 
 \hline
  \textbf{Case2}: \\
\hline
\textbf{Sentence 0}: The  Wigram  Baronetcy  of  Walthamstow  House  in  the  County  of   Essex is  a  title  in  the  Baronetage  of  the  \underline{\textbf{UNITED  KINGDOM}}. 
 \textbf{Sentence 3}:The  second  Baronet  also  represented  Wexford  Borough  in  \textbf{Parliament}.
\textbf{Sentence 5}: The  fourth  Baronet  was  a  Lieutenant  -  General  in  the  army  and  sat  as  a  Conservative  Member  of  \underline{\textbf{Parliament}} 
 for  South  Hampshire  and  Fareham \\
 \textbf{Correct answer}: {legislative\_body} \\
 \textbf{Baseline}: has\_part \\
 \textbf{DocRE-CLiP}:{legislative\_body} \\
 \hline
  \textbf{Case3}: \\
\hline
\textbf{Sentence 1}:[ Taking a more electronic music sound than his previous releases  TY.O was released in \underline{\textbf{December 2011}} by Universal Island Records but for  reasons unknown to Cruz  its British and American release were held off.]\textbf{ Sentence 3}:[ TY.O features a range of top - twenty and top - thirty singles including \enquote{Hangover} (featuring Florida) 
\textbf{\enquote{\underline{Troublemaker}}} \enquote{There She Goes} (sometimes featuring Pitbull)  the limited release \enquote{World in Our Hands} and \enquote{Fast Car} 
 which features on the Special Edition and Fast Hits versions of the album] \\
\textbf{Correct answer}: {Inception}  \\
\textbf{Baseline}: publication\_date \\
\textbf{DocRE-CLiP:} {publication\_date} \\


\hline
\end{tabular} 
}
\caption{Case study with DocRE-CLiP prediction. Underline text represents entities in the sentence, and purple color represents the DocRE-CLiP prediction.}
\label{casestudy}
\end{table}
\paragraph{\textbf{Effectiveness of link prediction model with context.}} Our focus has been on investigating link prediction models utilizing individual triples from the dataset. Throughout our analysis, we evaluate the performance of different link prediction models, specifically DistMult (Yang et al., 2015), Complex (Trouillon et al., 2016), R-GCN (Schlichtkrull et al., 2017), and KGE-HAKE (Zhang et al., 2022). Subsequently, we explored the influence of context on their performance. Notably, in each instance, upon the incorporation of context, the performance of the link prediction models improves.
Table~\ref{linkpredictionresults} summarizes the results. Considering these findings, RGCN has exhibited superior performance across metrics such as hits@1, hits@2, hits@10, and MRR. As a result, we have opted to select RGCN for link prediction.

\begin{table}[ht] 
 \centering

  \scalebox{0.7}{
\begin{tabular}{ccccc}

\hline

\textbf{Model} & \textbf{Metric} & \textbf{DocRED} & \textbf{ReDocRED} & \textbf{DWIE}\\
\hline
\multirow{3}{7em}{DistMult} & Hits@1 & 0.092 & 0.061 & 0.293  \\ 
& Hits@3 &  0.104 & 0.088& 0.307 \\ 
& Hits@10 &  0.127 & 0.111 & 0.334 \\ 
& MRR & 0.105 & 0.080 & 0.307 \\
\hdashline
\multirow{3}{7em}{DistMult+$_{context}$} & Hits@1 & 0.10 &  0.071 & 0.297 \\ 
& Hits@3 &  0.113 & 0.093 & 0.324 \\ 
& Hits@10 & 0.112  & 0.12 & 0.337 \\ 
& MRR & 0.11 & 0.12 & 0.43 \\
\hline

\multirow{3}{7em}{Complex} & Hits@1 & 0.092 & 0.076 & 0.286  \\ 
& Hits@3 & 0.097& 0.096& 0.296 \\ 
& Hits@10 & 0.110 &  0.104 & 0.317 \\ 
& MRR & 0.099 & 0.087 & 0.297 \\

\hdashline

\multirow{3}{7em}{Complex$_{context}$} & Hits@1 & 0.101 & 0.09  &  0.31\\ 
& Hits@3 & 0.12 &  0.10 & 0.33 \\ 
& Hits@10 &  0.15& 0.13  & 0.36\\ 
& MRR & 0.1 &  0.98 & 0.30\\

\hline

\multirow{3}{7em}{R-GCN} & Hits@1 & 0.06 & 0.033 & 0.38 \\ 
& Hits@3 & 0.09 & 0.06 & 0.43 \\ 
& Hits@10 & 0.11 & 0.091 & 0.45 \\ 
& MRR & 0.0827& 0.0532 & 0.40 \\

\hdashline
\multirow{3}{7em}{R-GCN$_{context}$} & Hits@1 &  0.11& 0.51  & 0.67  \\ 
& Hits@3 & 0.14 & 0.61 & 0.91\\ 
& Hits@10 &  0.23& 0.13 & 0.98\\ 
& MRR & 1.34 & 1.13 & 1.32\\

\hline
\multirow{3}{7em}{KGE-HAKE} & Hits@1 & 0.07 & 0.05  & 0.44  \\ 
& Hits@3 & 0.103 & 0.11& 0.45\\ 
& Hits@10 & 0.123  & 0.112 & 0.47\\ 
& MRR & 0.09 & 0.13 & 0.45\\
\hdashline
\multirow{3}{7em}{KGE-HAKE$_{context}$} & Hits@1 & 0.10  & 0.08 & 0.48  \\ 
& Hits@3 & 0.156 & 0.14 & 0.50  \\ 
& Hits@10 &  0.18 &  0.144 & 0.51\\ 
& MRR & 0.136 & 0.153 & 0.52 \\

\hline

\end{tabular} }
 \caption{{Performance of link prediction models.}}
\label{linkpredictionresults}
\end{table}

\begin{table}[ht]
\centering
\scalebox{0.75}{
\begin{tabular}{@{} p{1\linewidth}}
 \hline
 \textbf{Query}: (IBM research Brazil, parent\_organization, ?x)  \\
 \textbf{Answer}: IBM research  \\
 \textbf{Explanation}:   \\ \textit{IBM research Brazil, part\_of, IBM research} \\
 
 \hline

 \textbf{Query}: (Piraeus, country, ?x)  \\
 \textbf{Answer}: Greece   \\
 \textbf{Explanation}: \\
  \textit{\{Piraeus, located\_in\_the\_administrative\_territorial\_entity, Kiato\}} \\
  \textit{\{Kiato, country, Greece\}} \\
  \hline

  \textbf{Query}: (Quincy, country, ?x)  \\
 \textbf{Answer}: United States   \\
 \textbf{Explanation}:  \\
  \textit{\{Quincy, country, American\}} \\
  \textit{\{American, country\_of\_citizenship, America\}} \\  
  \textit{\{America, synonym, United States\}} \\  
 \hline

\end{tabular}
}
\caption{Example queries and results on DocRED dataset}
\label{pathresults}
\end{table}

\paragraph{\textbf{Study of the path for an explanation on DocRE-CLiP.}} Since explanation is an important part of our model, we analyze the performance of the traversal path. Table~\ref{pathresults} provides a few examples of explanation with respect to the hops for document-level relation extraction. This answers \textbf{RQ3}.

\section{Case Study}
We discuss two successful and one failed case of DocRE-CLiP and compare them with the baseline model, DRN (Table~\ref{casestudy}). 
\textbf{Case1}: To identify the relation between Manche and France in sentence 0, we use external knowledge about France and Mancy.
We get connecting context path using DocRE-CLiP: 
\textit{\{France, contains\_the\_administrative\_territorial\_entity, Normandy}\} 
\textit{\{Normandy, contains\_the\_administrative\_territorial\_entity, Manche}\}. 
Following the context path directly leads to relation \textit{contains\_the\_administrative\_territorial} between France and Mance. \textbf{Case2}:
With the aid of context path between entities,
\textit{\{United Kingdom, legislative\_body,	Parliament of the United Kingdom}\}  
\textit{\{Parliament of the United Kingdom,	instance\_of,	Parliament}\}. 
DocRE-CLiP successfully identifies correct relation \textit{legislative body}. \textbf{Case3}: Using pattern recognition, DocRE-CLiP identifies the publication date as a relation. However, the inclusion of the entity context, such as \enquote{Troublemaker}, \enquote{description}, and \enquote{song}, does not contribute significantly to the accuracy of the prediction. Consequently, DocRE-CLiP encounters difficulties in correctly predicting the relation.

\section{Conclusion}
This paper introduces DocRE-CLiP, a context-driven approach for document-level relation extraction (DocRE). Our results suggest that integrating diverse context types into the link prediction module enriches relation prediction within the DocRE framework providing interpretability. As future work, researchers can extend our efforts by crafting a versatile model capable of traversing diverse document types, thereby significantly amplifying its aptitude for assimilating knowledge. Furthermore, with conclusive evidence provided in our work as the first step, the document RE and KG link prediction research findings will interchangeably benefit. The code and the models are available at https://github.com/kracr/document-level-relation-extraction.

\section*{Acknowledgements}
We express our sincere gratitude to the Infosys Centre for Artificial Intelligence (CAI) at IIIT-Delhi for their  support. RK's effort has been supported by the U.S.~National Library of Medicine (through grant R01LM013240)


\bibliography{main}

\end{document}